\documentclass[twocolumn,showpacs,preprintnumbers,nofootinbib,amsmath,amssymb]{revtex4}

\usepackage{euscript,amssymb,amsmath}

\usepackage{graphicx}
\usepackage{epsfig}

\newcommand{\be}[1]{\begin{equation}\label{#1}}
\newcommand{\ee}{\end{equation}}
\newcommand{\ba}[1]{\begin{eqnarray}\label{#1}}
\newcommand{\ea}{\end{eqnarray}}
\newcommand{\rf}[1]{(\ref{#1})}
\newcommand{\nn}{\nonumber}

\begin{document}

\title{Asymptotic latent solitons, black strings and black branes in f(R)-gravity}

\author{Maxim Eingorn}\email{maxim.eingorn@gmail.com}  \author{Alexander Zhuk}\email{ai_zhuk2@rambler.ru}

\affiliation{Astronomical Observatory and Department of Theoretical Physics, Odessa National University, Street Dvoryanskaya 2, Odessa 65082, Ukraine}

\begin{abstract}
We investigate nonlinear $f(R)$ theories in the Kaluza-Klein models with toroidal compactification of extra dimensions. A point-like matter source has the dust-like
equation of state in our three dimensions and nonzero equations of state in the extra dimensions. We obtain solutions of linearized Einstein equations with this matter
source taking into account effects of nonlinearity of the model. There are two asymptotic regions where solutions satisfy the gravitational tests at the same level of
accuracy as General Relativity. According to these asymptotic regions, there are two classes of solutions. We call these solutions asymptotic latent solitons. The
asymptotic latent solitons from the first class generalize the known result of the linear theory. The asymptotic black strings and black branes are particular cases of
these asymptotic solutions. The second class of asymptotic solitons exists only in multidimensional nonlinear models. The main feature for both of these classes of
solutions is that the matter sources have tension in the extra dimensions.
\end{abstract}


\maketitle


\section{\label{sec:1}Introduction}

\setcounter{equation}{0}

Modern observational phenomena, such as dark energy and dark matter, are the great challenge  for present cosmology, astrophysics and theoretical physics. Within the
scope of the standard models, it has still not been offered a satisfactory explanation to these problems. This forces the search of solutions to these problems beyond
the standard models, for example, by considering modified gravitational theories. One of the possible generalizations consists in consideration of multidimensional
spacetimes. This generalization follows from the modern theories of unification such as superstrings, supergravity, and M-theory, which have the most self-consistent
formulation in spacetimes with extra dimensions \cite{Polchinski}. Nonlinear (with respect to the scalar curvature $R$) models $f(R)$ are the other possible
generalization of gravitational theories. Nonlinear models may arise either due to quantum fluctuations of matter fields including gravity \cite{BirrDav}, or as a result
of compactification of extra spatial dimensions \cite{NOcompact}. Starting from the pioneering paper \cite{Star1}, the nonlinear theories of gravity $f(R)$ have
attracted the great deal of interest because these models can provide a natural mechanism of the early inflation. Recently, it was realized that these models can also
explain the late-time acceleration of the Universe. This fact resulted in a new wave of papers devoted to this topic (see, e.g., reviews \cite{review1}-\cite{review4}).
It seems natural to combine these two approaches, i.e. to consider multidimensional nonlinear models. Different aspects of these models (e.g., the early inflation and
the late-time acceleration) were considered in a number of papers (see, e.g., \cite{GMZ1,GMZ2}).

Obviously, these physical theories should be consistent with observations. Our previous paper \cite{EZf(R)} was devoted to this important problem. We have considered a
point-like massive source in nonlinear $f(R)$ models with toroidal compactification of the extra dimensions. Such massive source with dust-like equation of state is a
good approach to calculate the well known gravitational tests (perihelion shift, deflection of light and time delay of radar echoes) in General Relativity. So, it seems
natural to generalize it to the multidimensional case and also suppose the dust-like equations of state in all spatial dimensions. However, we have shown that such
matter source contradicts the observations in the case of extra dimensions. The similar situation takes place in multidimensional linear models \cite{EZ5,EZ4}, where a
point-like mass with dust-like equations of state in all spatial dimensions also contradicts the experimental data. However, latent solitons, in particular, black
strings and black branes, satisfy the gravitational experiments at the same level of accuracy as General Relativity. A distinctive feature of these solutions is that the
matter source has tension in the internal space. Therefore, in nonlinear models, it also makes sense to suppose that the matter source has nonzero equations of state in
the extra dimensions.

Thus, in our present paper we consider the matter source with the dust-like equation of state in our three dimensions but with nonzero equations of state in the extra
dimensions. We get solutions of linearized Einstein equations with this matter source taking into account effects of nonlinearity of the model. We show that there are
two classes of solutions which satisfy the gravitational tests in asymptotic regions at the same level of accuracy as General Relativity. We call these solutions
asymptotic latent solitons. The asymptotic latent solitons from the first class generalize the known results \cite{EZ5,EZ4} of the linear theory. The second class of
asymptotic solitons exists only in multidimensional nonlinear models. The main feature for both of these classes of solutions is that the matter sources have tension in
the extra dimensions.

The paper is structured as follows. In Sec. II, we consider a weak-field limit of $f(R)$ theories. Here, a point-like matter source has nonzero equations of state in the
extra dimensions. We obtain solutions of linearized Einstein equations with this matter.  In Sec. III, we find conditions for asymptotic latent solitons. These solutions
satisfy the gravitational experiments similar to General Relativity. The main results are summarized in the concluding Sec. IV.


\section{\label{sec:2}Approximate solutions}

Equations of motion for $f(R)$ gravitational theories in the case of an arbitrary number of spacetime dimensions $\mathcal{D}=1+D\geq 4$ are given in many papers (see,
e.g., \cite{GMZ1,GMZ2}). For our purpose, we shall follow equations in our previous paper \cite{EZf(R)}. Saying "approximate solutions" we mean that we are looking for
solutions of $f(R)$ gravitational theories in the weak field approximation.  This means that the gravitational field is weak and the velocities of test bodies are small
compared to the speed of light $c$. In this case the metrics is only slightly perturbed from its flat spacetime value:
\be{2.1} g_{ik}\approx\eta_{ik}+h_{ik}\, , \ee
where $h_{ik}$ are corrections of the order $1/c^2$. We also suppose that $f(R)$ is an analytical function and can be expanded in a Taylor series near the flat spacetime
background value $R=0$ :
\be{2.2} f(R)=R+aR^2+o\left(R^2\right),\quad |R|\gg|a|R^2\, , \ee
where $o\left(R^n\right)$ means that this function contains the terms $R^m$ with powers $m>n$. In Eq. \rf{2.2}, $a\equiv (1/2)f''(0)$ and we assumed that $f(0)=0$. The
latter means that the cosmological constant $\Lambda$ is absent in the model in accordance with our assumption that the background spacetime is flat\footnote{The
generalization to the case of the de Sitter background with metric coefficients $g_{0ik}$ and the scalar curvature $R_0$ is straightforward. In all our concluding
formulas we should make the substitution $f''(0)\to f''(R_0)$ (see \cite{EZf(R)}).}. We also normalize our model in such a way that $f'(0)=1$ which provides the
transition to the usual form of the Einstein equations in the limit $R \to 0 \; \Rightarrow \; f(R) \to R$.

Following the paper \cite{EZf(R)}, we arrive at the system of equations
\be{2.3}
\triangle_3\left(h_{ik}+\frac{4a}{D-1}\eta_{ik}R\right)\approx\frac{4S_D\tilde{G}_{\mathcal D}}{c^4}\left(T_{ik}-\frac{1}{D-1}T\eta_{ik}\right)
\ee
and
\be{2.4}
\triangle_3 R+\frac{D-1}{4aD}R\approx-\frac{S_D\tilde{G}_{\mathcal D}}{aDc^4}T\, ,
\ee
where $S_D=2\pi^{D/2}/\Gamma (D/2)$ is the total solid angle (the surface area of the $(D-1)$-dimensional sphere of a unit radius), $\tilde G_{\mathcal{D}}$ is the
gravitational constant in $\mathcal{D}$-dimensional spacetime and $\triangle_3=\delta^{\alpha\beta}\cfrac{\partial^2}{\partial x^{\alpha}\partial x^{\beta}}\,
,\alpha,\beta =1,2,3$ is the $3$-dimensional Laplace operator. In these equations, the energy-momentum tensor is not defined yet. We only require that $T_{ik}=O(c^2)$
and the matter source is uniformly smeared over the internal space. Let us suppose that the nonzero components of the energy-momentum tensor read
\ba{2.5}
T_{00}&\approx&\frac{1}{V_d}\rho_3({\bf r}_3)c^2\, ,\nn \\
T_{\mu\mu}&\approx&\omega_{\mu}\frac{1}{V_d}\rho_3({\bf r}_3)c^2\approx\omega_{\mu}T_{00},\quad
\mu=4,5,\ldots,D\, ,
\ea
where $\rho_3({\bf r}_3)$ is the three-dimensional rest mass density, ${\bf r}_3$ is the three-dimensional radius vector, $V_d$ is the volume of the
$(d=D-3)$-dimensional internal space and $\omega_{\mu}$, $\mu=4,5,\ldots,D$ are arbitrary constants which define equations of state in the extra dimensions. In the case
of a point-like mass with dust-like equations of state in all spatial dimensions all $\omega_{\mu}=0$ and this is the case of our previous paper \cite{EZf(R)}.

For the energy-momentum tensor \rf{2.5} the Eq. \rf{2.3} takes the form
\be{2.6}
\triangle_3\left(h_{00}+\frac{4a}{D-1}R\right)\approx\frac{4S_D\tilde{G}_{\mathcal D}}{c^2}\frac{1}{V_d}\rho_3\left(\frac{D-2+\Omega}{D-1}\right),
\ee
\ba{2.7}
\triangle_3\left(h_{\beta\beta}-\frac{4a}{D-1}R\right)&\approx&\frac{4S_D\tilde{G}_{\mathcal D}}{c^2}\frac{1}{V_d}\rho_3\left(\frac{1-\Omega}{D-1}\right),\nn\\
\beta&=&1,2,3
\ea
and
\ba{2.8}
\triangle_3\left(h_{\mu\mu}-\frac{4a}{D-1}R\right)&\approx&\frac{4S_D\tilde{G}_{\mathcal D}}{c^2}\frac{1}{V_d}\rho_3\left(\omega_{\mu}+\frac{1-\Omega}{D-1}\right),\nn\\
\mu&=&4,5,\ldots,D\, ,
\ea
where $\Omega=\sum\limits_{\mu=4}^D\omega_{\mu}$. For the Eq. \rf{2.4} we get
\be{2.9}
\triangle_3 R+\frac{D-1}{4aD}R\approx-\frac{S_D\tilde{G}_{\mathcal D}}{aDc^2}(1-\Omega)\cfrac{1}{V_d}\rho_3({\bf r}_3)\, .
\ee

In the equations above, we did not use spherical symmetry of the matter source. Let us consider now the important case of a point-like mass with $\rho_3({\bf
r}_3)=m\delta({\bf r}_3)$. This enables us to solve the Eqs. \rf{2.6}-\rf{2.9}. First, we consider the case $a<0$. If $a=0$, then we get a linear model that is not of
interest for us. The case $a>0$ will be briefly discussed later. So, for $a<0$ we find:
\be{2.10}
h_{00}\approx-\frac{4a}{D-1}R-\frac{S_D\tilde{G}_{\mathcal D}}{c^2\pi V_d}\left(\frac{D-2+\Omega}{D-1}\right)\frac{m}{r_3}\, ,
\ee
\be{2.11}
h_{\beta\beta}\approx\frac{4a}{D-1}R-\frac{S_D\tilde{G}_{\mathcal D}}{c^2\pi V_d}\left(\frac{1-\Omega}{D-1}\right)\frac{m}{r_3},\quad\beta=1,2,3\, ,
\ee
\ba{2.12}
h_{\mu\mu}&\approx&\frac{4a}{D-1}R-\frac{S_D\tilde{G}_{\mathcal D}}{c^2\pi V_d}\left(\omega_{\mu}+\frac{1-\Omega}{D-1}\right)\frac{m}{r_3}\, ,\nn\\
\mu&=&4,5,\ldots,D
\ea
and
\be{2.13}
R\approx\frac{S_D\tilde{G}_{\mathcal D}(1-\Omega)}{4aDc^2\pi V_d}\cfrac{m}{r_3}\exp\left[-\left(\frac{D-1}{4|a|D}\right)^{1/2}r_3\right]\, .
\ee
To investigate the concordance with gravitational experiments, we need to compare $00$ and $\beta\beta$ components of the metric perturbations. Substituting \rf{2.13}
into \rf{2.10} and \rf{2.11}, we obtain
\ba{2.14}
h_{00}&=&-\frac{S_D\tilde{G}_{\mathcal D}}{(D-1)c^2\pi
V_d}\frac{m}{r_3}\left\{D-2+\Omega\phantom{\int}\right.\nn\\
&+&\left.\frac{1-\Omega}{D}\exp\left[-\left(\frac{D-1}{4|a|D}\right)^{1/2}r_3\right]\right\}
\ea
and
\ba{2.15}
h_{\beta\beta}&=&-\frac{S_D\tilde{G}_{\mathcal D}}{(D-1)c^2\pi
V_d}\frac{m}{r_3}\left\{1-\Omega\phantom{\int}\right.\nn \\
&-&\left.\frac{1-\Omega}{D}\exp\left[-\left(\frac{D-1}{4|a|D}\right)^{1/2}r_3\right]\right\}\, ,\nn \\
\beta&=&1,2,3\, .
\ea


\section{\label{sec:3}Asymptotic latent solitons, black strings and black branes}

It is well known that to satisfy the gravitational experiments (the deflection of light, the time delay of radar echoes) at the same level of accuracy as General
Relativity, the metric coefficients $h_{00}$ and $h_{\beta\beta}$ should coincide with each other. For our model, we can easily achieve it for two distinctive asymptotic
regions: $r_3\gg|a|^{1/2}$ and  $r_3\ll|a|^{1/2}$. Strictly speaking, we do not know the value of $a$. Therefore, we can find ourselves in any of these regions. We
require only that our model should be consistent with the experimental data in the asymptotic region under consideration. It is worth recalling that we consider the case
of a weak nonlinearity $|R|\gg |a|R^2$ (see \rf{2.2}). Let us consider these two regions separately.

\subsection{$r_3\gg\sqrt{|a|}$}

If this condition is satisfied on a small terrestrial scale, it is clear that it will run on a large astrophysical scale. In this case, we can drop the exponents in Eqs.
\rf{2.14} and \rf{2.15}. It means that the effect of nonlinearity is negligible in this region and for the metric coefficients we obtain
\be{3.1}
h_{00}=-\frac{S_D\tilde{G}_{\mathcal D}}{(D-1)c^2\pi V_d}\frac{m}{r_3}(D-2+\Omega)\, ,
\ee
\be{3.2}
h_{\beta\beta}=-\frac{S_D\tilde{G}_{\mathcal D}}{(D-1)c^2\pi V_d}\frac{m}{r_3}(1-\Omega),\quad\beta=1,2,3\, .
\ee
Now, equating $h_{00}$ and $h_{\beta\beta}$, we arrive at the following equation:
\be{3.3}
\Omega=\sum\limits_{\mu=4}^D\omega_{\mu}=-\frac{D-3}{2}=-\frac{d}{2}\,
\ee
in full analogy with the condition (31) for latent solitons in \cite{EZ5}, where the parameter $\gamma_{\mu}=2\omega_{\mu}+1$ and all $d_{\mu}=1$. Then, we obtain from
\rf{3.3} the latent soliton condition: $\sum_{\mu=4}^D\gamma_{\mu}=0$. Therefore, the condition \rf{3.3} defines asymptotic latent solitons in $f(R)$-gravity. In a
particular case $D=4 \to d=1$, we obtain asymptotic black strings with the equation of state $\omega=-1/2$ in the extra dimension. For asymptotic black branes with
$D>4$, we have the same equation of state $\omega_{\mu}=-1/2$ in all extra dimensions. Therefore, all these asymptotic solutions have tension in the extra dimensions. Similar to the linear model, the extra dimensions are flat ($h_{\mu\mu}=0,\, \mu =4,5,\ldots,D$) for the asymptotic black strings and black branes.
The case $D=3$ corresponds to a usual point-like mass in the three-dimensional space. Obviously, here we have transition to General Relativity.

It is natural to suppose that in our asymptotic region
\be{3.4}
h_{00}=\frac{2\varphi_N}{c^2},\quad\varphi_N=-\frac{G_Nm}{r_3}\, .
\ee
Then, we obtain from \rf{3.1} and \rf{3.3} the relation between the Newton's and multidimensional gravitational constants:
\be{3.5}
G_N=\frac{S_D\tilde{G}_{\mathcal D}}{4\pi V_d}\, .
\ee

As we mentioned above, the expansion \rf{2.2} works if $|aR|\ll 1$. From the Eqs. \rf{2.13}, \rf{3.3} and \rf{3.5} we get
\be{3.6}
aR\approx\frac{D-1}{2D}\cfrac{G_Nm}{c^2r_3}\exp\left[-\left(\frac{D-1}{4|a|D}\right)^{1/2}r_3\right]\, .
\ee
which is really much less than 1. First, it follows due to the smallness of the exponent and, second, due to the weak-field condition $|\varphi_N/c^2|\ll 1$.

\subsection{$r_3\ll\sqrt{|a|}$}

Obviously, if this condition is satisfied on an astrophysical scale, then it will work also for a terrestrial scale (e.g., for the table-top inverse square law
experiments). In this case, we can replace the exponents in \rf{2.14} and \rf{2.15} by unities. Therefore, the effect of nonlinearity is not negligible in this region,
although the weak nonlinearity condition $|aR|\ll 1$ should be satisfied. For the metric coefficients we obtain
\be{3.7} h_{00}=-\frac{S_D\tilde{G}_{\mathcal D}}{(D-1)c^2\pi V_d}\frac{m}{r_3}\left(D-2+\Omega+\frac{1-\Omega}{D}\right)\, , \ee
\be{3.8} h_{\beta\beta}=-\frac{S_D\tilde{G}_{\mathcal D}}{(D-1)c^2\pi V_d}\frac{m}{r_3}\left(1-\Omega-\frac{1-\Omega}{D}\right),\ \beta=1,2,3\, . \ee
Equating these expressions to each other, we obtain the following condition of agreement with gravitational tests:
\be{3.9}
\Omega=\sum\limits_{\mu=4}^D\omega_{\mu}=-\frac{D-2}{2}=-\frac{d+1}{2}\, .
\ee
Similarly to the previous case, the asymptotic solutions obeying this condition satisfy the gravitational tests with the same accuracy as General Relativity. Therefore,
we find the second class of asymptotic latent solitons. It can be easily seen that there is no transition to three-dimensional case. Thus, this class of latent solitons
exists only in the multidimensional space ($D>3$). Because $\Omega <0 $, these latent solitons also have tension in the extra dimensions. It can be easily seen that for
the value of $\Omega$ from the Eq. \rf{3.9} there are no asymptotic solutions with $h_{\mu\mu}=0,\, \mu =4,5,\ldots,D$, i.e. with flat extra dimensions.

If we suppose for the considered region the natural condition \rf{3.4}, then we easily get (with the help of the Eqs. \rf{3.7} and \rf{3.9}) the relation \rf{3.5} for
the Newton's and multidimensional gravitational constants. Taking into account equations \rf{2.13}, \rf{3.5} and \rf{3.9}, we also find that the weak nonlinearity
condition
\be{3.10}
aR\approx\frac{1}{2}\cfrac{G_Nm}{c^2r_3} \ll 1
\ee
is satisfied due to the weak-field condition $|\varphi_N/c^2|\ll 1$.

To conclude this section, we want to discuss briefly the case $a>0$. For such $a$, the exponents in Eqs. \rf{2.13}-\rf{2.15} should be replaced by cosines. Obviously, it
makes sense to consider only the asymptotic region $r_3\ll |a|^{1/2}$ (because solutions with cosines are nonphysical in the opposite case). Then, we exactly reproduce
the previous case B.

\section{Conclusion}

In nonlinear $f(R)$ multidimensional models with toroidal compactification of the extra dimensions, we searched for solutions which satisfy the classical gravitational
tests (deflection of light and time delay of radar echoes). For an arbitrary function $f(R)$, it is hardly possible to solve equations of motion exactly. Therefore, we
have investigated these equations in a weak-field limit when the metric tensor can be split into a flat background and a weak perturbation. In this approximation,
nonlinearity is also weak and analytical function $f(R)$ can be expanded in a Taylor series: $f(R)=R+aR^2+o\left(R^2\right) ,\quad a\equiv (1/2)f''(0)$. We have shown in
our previous paper \cite{EZf(R)} that the most natural point-like massive source with the dust-like equations of state in all dimensions contradicts the experimental
data.
The similar situation takes place in multidimensional linear models \cite{EZ5,EZ4}. In such linear models, latent solitons, in particular, black strings and black
branes, satisfy the gravitational experiments at the same level of accuracy as General Relativity. A distinctive feature of these solutions is that the matter source has
tension in the internal space. Therefore, in the present paper we also supposed that a matter source is still dust-like in our three dimensions but has nonzero equations
of state in the extra dimensions. Then, we have found solutions of linearized Einstein equations with this matter source taking into account effects of nonlinearity of
the model. These solutions demonstrate explicitly that there are two asymptotic regions where our model is in agreement with the gravitational tests similar to General
Relativity. In the first asymptotic region $r_3 \gg |a|^{1/2}$, the effect of nonlinearity is negligible and we found the condition for solutions which exactly coincides
with the condition for latent solitons in the linear model \cite{EZ5}. This is a natural result and we called these solutions asymptotic latent solitons (asymptotic
black strings and asymptotic black branes are particular cases of these solutions). In the second asymptotic region $r_3 \ll |a|^{1/2}$, the effect of nonlinearity plays
an important role. As a result, we got here a new condition for solutions to satisfy the gravitational tests. This is the second class of asymptotic latent solitons. We
have shown that these solitons exist only in the multidimensional case while for the first class of asymptotic latent solitons there is the transition to the
three-dimensional case. Therefore, we have shown that in nonlinear multidimensional models there are solutions which satisfy the gravitational tests. The main feature
for both of these classes of solutions is that the matter sources have tension in the extra dimensions.

\section*{ACKNOWLEDGEMENTS}

This work was supported in part by the "Cosmomicrophysics" programme of the Physics and Astronomy Division of the National Academy of Sciences of Ukraine. A.~Zh.
acknowledges the hospitality of the Theory Division of CERN during the final preparation of this work.


\end{document}